\documentclass{emulateapj}

\newcommand{\hhh}{H$_3^+$}
\newcommand{\hh}{H$_2$}
\newcommand{\ohhh}{\textit{o}-\hhh}
\newcommand{\phhh}{\textit{p}-\hhh}
\newcommand{\ohh}{\textit{o}-\hh}
\newcommand{\phh}{\textit{p}-\hh}
\newcommand{\ortho}{\textit{ortho}}
\newcommand{\para}{\textit{para}}

\usepackage{amssymb}

\begin{document}

\title{On the \textit{Ortho}:\textit{Para} Ratio of \hhh\ in Diffuse Molecular Clouds\altaffilmark{1}}

\author{Kyle N. Crabtree\altaffilmark{2},
Nick Indriolo\altaffilmark{3},
Holger Kreckel\altaffilmark{2},
Brian A. Tom\altaffilmark{2,4},
Benjamin J. McCall\altaffilmark{2,3,5}
}
\altaffiltext{1}{Based in part on observations made with ESO Telescopes at the La Silla or Paranal Observatories under programme ID 384.C-0618}
\altaffiltext{2}{Department of Chemistry, University of Illinois at Urbana-Champaign, Urbana, IL 61801}
\altaffiltext{3}{Department of Astronomy, University of Illinois at Urbana-Champaign, Urbana, IL 61801}
\altaffiltext{4}{Present Address: Department of Chemistry, United States Air Force Academy, CO 80840, USA}
\altaffiltext{5}{Department of Physics, University of Illinois at Urbana-Champaign, Urbana, IL 61801}

\begin{abstract}
The excitation temperature $T_{01}$ derived from the relative intensities of the $J=0$ (\para) and $J=1$ (\ortho) rotational levels of \hh\ has been assumed to be an accurate measure of the kinetic temperature in interstellar environments. In diffuse molecular clouds, the average value of $T_{01}$ is $\sim$70 K. However, the excitation temperature $T$(\hhh) derived from the $(J,K) = (1,1)$ (\para) and $(1,0)$ (\ortho) rotational levels of \hhh\ has been observed to be $\sim$30 K in the same types of environments. In this work, we present observations of \hhh\ in three additional diffuse cloud sight lines for which \hh\ measurements are available, showing that in 4 of 5 cases $T_{01}$ and $T$(\hhh) are discrepant. We then examine the thermalization mechanisms for the \ortho:\para\ ratios of \hhh\ and \hh, concluding that indeed $T_{01}$ is an accurate measure of the cloud kinetic temperature, while the \ortho:\para\ ratio of \hhh\ need not be thermal. By constructing a steady-state chemical model taking into account the nuclear-spin-dependence of reactions involving \hhh, we show that the \ortho:\para\ ratio of \hhh\ in diffuse molecular clouds is likely governed by a competition between dissociative recombination with electrons and thermalization via reactive collisions with \hh.
\end{abstract}


\section{INTRODUCTION}

Observations of H$_3^+$ in diffuse molecular clouds \citep[diffuse clouds in which a significant fraction of the hydrogen is in molecular form;][]{snow2006} have led to various unexpected discoveries.  The very first detection of H$_3^+$ along a diffuse molecular cloud sight line (toward Cyg OB2 12) showed an abundance about 10 times greater than expected for the environment \citep{mccall1998a}.  This surprising overabundance---also found toward several more diffuse cloud sight lines---led to the eventual conclusion that the ionization rate of H$_2$ due to cosmic rays, $\zeta_2$, must be about 1 order of magnitude larger than previously thought \citep{mccall2003,indriolo2007}.  Another puzzling outcome from H$_3^+$ observations is that the average excitation temperature derived from the two lowest energy states \citep[$\langle T({\rm H}_3^+)\rangle\approx30$~K;][]{indriolo2007} differs significantly from the average excitation temperature derived from the two lowest rotational states of H$_2$ \citep[$\langle T_{01}\rangle\approx70$~K;][]{savage1977,rachford2002,rachford2009}.  Given that both species are expected to be thermalized to the cloud kinetic temperature by collisions, such a discrepancy is unexpected.

Despite the fact that the $J=0$ and $J=1$ levels of H$_2$ are essentially different ``species''---conversion between the even--$J$ {\it para} and odd--$J$ {\it ortho} nuclear spin states is only possible through reactive collisions---it has long been assumed that $T_{01}$ is a good approximation for the kinetic temperature, $T_{kin}$, in diffuse molecular clouds.  This is because H$_2$ molecules are expected to experience many reactive collisions with protons during their lifetimes.  As a result, the {\it ortho} and {\it para} populations of H$_2$ should be brought into thermal equilibrium with the proton kinetic temperature \citep{dalgarno1973}.

Similarly, H$_3^+$ also has {\it ortho} and {\it para} nuclear spin states ($(J,K)=(1,1)$ is the lowest lying {\it para} state, and $(J,K)=(1,0)$ the lowest lying {\it ortho} state).  For H$_3^+$ though, the relative population between the two states has been expected to be thermalized by reactive collisions with H$_2$ \citep{mccall1998b,mccall2003,gibb2010}.  As the average values of $T_{01}$ and $T({\rm H}_3^+)$ do not agree, this does not seem to be the case.

However, out of the 66 and 18 sight lines used to compute $\langle T_{01}\rangle$ and $\langle T({\rm H}_3^+)\rangle$, respectively, only 2 are shared between both samples.  While previously reported values of $T_{01}$ and $T({\rm H}_3^+)$ do differ for these sight lines toward $\zeta$~Per and X~Per by about 30~K \citep{savage1977,rachford2002,indriolo2007}, such a small sample does not provide particularly meaningful results.  In order to increase the number of sight lines with {\it ortho} and {\it para} column densities determined for both H$_2$ and H$_3^+$, we have made observations searching for H$_3^+$ absorption features along sight lines with measured H$_2$ column densities.  These observations and our data reduction procedures are described in Section 2.  Section 3 discusses and justifies many of the assumptions made concerning molecular hydrogen and $T_{01}$ in diffuse molecular clouds. In Section 4, we examine the \hhh\ + \hh\ reaction and present steady state models to explore the temperature discrepancy between $T_{01}$ and $T$(\hhh). Section 5 reviews the observations and modeling, and presents our conclusions from the work.

\section{OBSERVATIONS \& DATA REDUCTION}

Target sight lines were selected based on H$_2$ column densities \citep{savage1977,rachford2002,rachford2009} and $L$-band magnitudes.  We required that the $J=0$ and $J=1$ column densities of H$_2$ ($N(0)$ and $N(1)$, respectively) both be known, and that the total H$_2$ column density ($N({\rm H}_2)$) be greater than $10^{20}$~cm$^{-2}$.  The $L$-band magnitude was required to be brighter than 6~mag so that the necessary integration times would be relatively short.  Observations focused on the $R(1,1)^u$, $R(1,0)$, and $R(1,1)^l$ transitions (at 3.668083~$\mu$m, 3.668516~$\mu$m, and 3.715479~$\mu$m, respectively) which arise from the $(J,K)=(1,1)$ and $(1,0)$ levels of the ground vibrational state of H$_3^+$, the only levels expected to be significantly populated at average diffuse cloud temperatures ($T\sim70$~K).

Spectra in support of this project were obtained using the Phoenix spectrometer \citep{hinkle2003} at the Gemini South Telescope and the Cryogenic High-resolution Infrared Echelle Spectrograph (CRIRES)~\citep{kaufl2004} at the Very Large Telescope (VLT).  Observations at Gemini South were made in queue mode, and the Phoenix spectrometer was used with its echelle grating and 0.17'' slit to produce a resolving power of about 70,000, and with the L2734 filter to select the order containing the $R(1,1)^u$ and $R(1,0)$ transitions.  Observations at the VLT were made in service mode, and CRIRES was used with its 0.2'' slit to provide a resolving power of about 100,000, and a reference wavelength of 3715.0~nm to cover the $R(1,1)^u$ and $R(1,0)$ transitions on detector 1 and the $R(1,1)^l$ transition on detector 3.  The adaptive optics system was used with CRIRES to maximize starlight passing through the narrow slit.  In addition to the science targets, bright, early-type stars were observed for use as telluric standards.  For all observations, the star was nodded along the slit in an ABBA pattern in order to facilitate the removal of atmospheric emission lines and dark current via the subtraction of neighboring images.  A log containing the observed sight lines and respective integration times is given in Table \ref{tblobs}.

\begin{deluxetable*}{llccc}
\tablecaption{Observations \label{tblobs}}
\tablehead{
 & & & \colhead{Integration Time} & \\
\colhead{Object} & \colhead{Date(s) of Observation} & \colhead{Telescope} & \colhead{(min)} & \colhead{Standard}
}
\startdata
HD 149404  & 2009 Aug 05 & Gemini South & 32 & $\lambda$ Sco \\
$\chi$ Oph & 2009 Aug 30 & Gemini South & 12 & $\lambda$ Sco \\
HD 152236  & 2009 Aug 30 & Gemini South & 16 & $\lambda$ Sco \\
HD 154368  & 2009 Aug 30 & Gemini South & 32 & $\lambda$ Sco \\
HD 53367   & 2009 Dec 03 & VLT          & 30 & $\kappa$ Ori \\
HD 73882   & 2009 Dec 03 & VLT          & 30 & $\zeta$ Pup \\
           & 2010 Jan 17 & VLT          & 30 & $\zeta$ Pup  \\
HD 110432  & 2010 Jan 27 & VLT          & 10 & $\eta$ Cen \\
           & 2010 Feb 28 & VLT          & 20 & $\eta$ Cen \\
           & 2010 Mar 02 & VLT          & 40 & $\eta$ Cen \\
$\mu$ Nor  & 2010 Apr 05 & Gemini South & 72 & $\lambda$ Sco
\enddata
\end{deluxetable*}

Various standard IRAF\footnote{http://iraf.noao.edu/} procedures were used in the data reduction process.  Given the different state of data available from Phoenix versus CRIRES, different amounts of processing were required for data from the 2 telescopes.  For each night of Phoenix observations, a bad pixel map was created from the average of several dark frames, and these pixels were interpolated over in the object and flat frames.  Flats were then combined to create a normalized flat field which was divided out of the object frames.  Neighboring AB image pairs were subtracted from each other to remove atmospheric emission and dark current.  Finally, one-dimensional spectra were extracted using {\it apall}.  In the case of CRIRES observations, data were processed through the CRIRES specific pipeline, resulting in fully reduced two-dimensional spectral images for each target (given the product codes SCOM and PCOM by the pipeline).  One-dimensional spectra were extracted from these images.  All such spectra from both telescopes were then imported to IGOR Pro\footnote{http://www.wavemetrics.com/} where the remainder of reduction was performed.

\begin{figure}
\epsscale{1.0}
\plotone{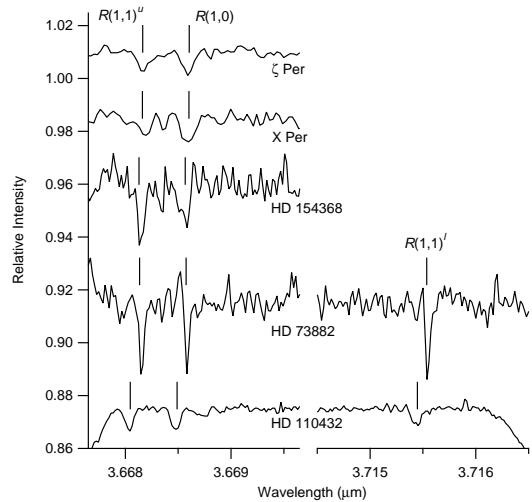}
\caption{Spectra showing absorption lines from the $(J,K)=(1,1)$ and $(1,0)$ states of H$_3^+$.  Spectra for $\zeta$~Per and X~Per were taken at UKIRT and are reprocessed versions of the data previously reported in \citet{indriolo2007}.  The spectrum for HD 154368 was taken at Gemini South, while those for HD 73882 and HD 110432 were taken at the VLT.  Large deviations from flat continuum levels in the spectra for HD 154368, HD 73882, and HD 110432 are the combination of 2 effects: (1) artifacts due to continuum fitting at wavelengths far away from the H$_3^+$ absorption lines; (2) inability to remove the strong atmospheric methane line immediately shortward of the $R(1,1)^u$ line.  Vertical lines mark the expected positions of the H$_3^+$ absorption lines given previously determined interstellar gas velocities along each sight line.}
\label{figspectra}
\end{figure}

Individual Phoenix spectra within an exposure sequence for a given target were then added together.  Science target spectra from both telescopes were divided by telluric standard spectra to remove atmospheric absorption features and to normalize each spectrum.  These normalized spectra were wavelength calibrated with a typical accuracy of $\sim2$~km~s$^{-1}$ using the vacuum wavelengths of the atmospheric absorption lines.  Calibrated spectra were then shifted into the local standard of rest (LSR) frame, and spectra of the same target from different nights were combined using a variance-weighted mean.  For each sight line the continuum surrounding the absorption features was then fit with an $n^{\rm th}$ order polynomial ($n=7-9$) and the spectrum was divided by the fit to re-normalize the continuum level.  The resulting spectra for sight lines with H$_3^+$ absorption features---HD 154368, HD 73882, and HD 110432---are shown in Figure \ref{figspectra}.  Although H$_3^+$ spectra for $\zeta$~Per and X~Per (the other 2 sight lines with both H$_2$ and H$_3^+$ data) are reported in \citet{indriolo2007}, the reduction process utilized in that study did not combine spectra using a variance-weighted mean, nor did it fit fluctuations in the continuum level with a polynomial function.  For the purpose of consistency, we have reprocessed the data from both sight lines.  No new data have been added, but differences in the method of processing have resulted in output spectra with slightly better signal-to-noise ratios (S/N).  These spectra are also shown in Figure \ref{figspectra}.

Absorption features due to H$_3^+$ were fit with Gaussian functions in order to determine equivalent widths, velocity full-width at half-maxima (FWHM), and interstellar gas velocities.  Our fitting procedure uses the functional form of a Gaussian where area (as opposed to amplitude) is a free parameter, and includes a fit to the continuum level, $y_0$.  In the case of the $R(1,1)^u$ and $R(1,0)$ lines, both absorption features are fit simultaneously and a single best-fit continuum level is found.  Uncertainties on the equivalent widths ($\delta W_{\lambda}$) and continuum level ($\delta y$)---both at the 1$\sigma$ level---were output by the fitting process.  To estimate the systematic uncertainties due to continuum placement, we forced the continuum level to $y_0+\delta y$ and $y_0-\delta y$ and re-fit the absorption lines.  Variations in the equivalent widths due to this shift are small compared to those reported by the fitting procedure and so are not included in our analysis (i.e., $\sigma(W_{\lambda})$=$\delta W_{\lambda}$).
%Uncertainties were computed from the standard deviation, $\sigma$, of the residuals after subtracting the Gaussian fits, using the formula $\sigma(W_\lambda) = \sigma\sqrt{\mathcal{N}}\Delta\lambda$, where $\mathcal{N}$ is the number of pixels spanned by the absorption feature and $\Delta\lambda$ is the wavelength coverage of a single pixel.
Assuming optically thin absorption lines and taking transition dipole moments and wavelengths from \citet{goto2002} and references therein, column densities were derived from equivalent widths using the standard relation.  All of these results are shown in Table \ref{tbllineparam}.

\begin{deluxetable*}{llcccccc}
\tablecaption{Absorption Line Parameters \label{tbllineparam}}
\tablehead{
 & & \colhead{$v_{\rm LSR}$} & \colhead{FWHM} & \colhead{$W_{\lambda}$} & \colhead{$\sigma(W_{\lambda})$} &
\colhead{$N(J,K)$} & \colhead{$\sigma(N)$} \\
\colhead{Object} & \colhead{Transition} & \colhead{(km s$^{-1}$)} & \colhead{(km s$^{-1}$)} &
\colhead{($10^{-6}~\mu$m)} & \colhead{($10^{-6}~\mu$m)} & \colhead{($10^{13}$ cm$^{-2}$)} & \colhead{($10^{13}$ cm$^{-2}$)}
}
\startdata
$\zeta$ Per& $R(1,1)^u$ &  7.7 &11.0 & 0.99 & 0.13 & 4.09 & 0.53 \\
           & $R(1,0)$   &  6.1 & 9.0 & 1.00 & 0.11 & 2.53 & 0.29 \\
X Per      & $R(1,1)^u$ &  8.2 & 9.1 & 0.80 & 0.17 & 3.34 & 0.69 \\
           & $R(1,0)$   &  6.3 &10.2 & 1.30 & 0.18 & 3.29 & 0.45 \\
HD 154368  & $R(1,1)^u$ &  5.4 & 6.0 & 1.79 & 0.30 & 7.43 & 1.24 \\
           & $R(1,0)$   &  5.2 & 5.6 & 1.12 & 0.29 & 2.83 & 0.74 \\
HD 73882   & $R(1,1)^u$ &  5.9 & 3.9 & 1.44 & 0.21 & 5.97 & 0.86 \\
           & $R(1,0)$   &  5.7 & 3.2 & 1.16 & 0.19 & 2.94 & 0.48 \\
           & $R(1,1)^l$ &  5.4 & 3.5 & 1.34 & 0.15 & 6.15 & 0.69 \\
HD 110432  & $R(1,1)^u$ & -3.8 & 6.9 & 0.74 & 0.06 & 3.08 & 0.24 \\
           & $R(1,0)$   & -3.3 & 7.5 & 0.83 & 0.07 & 2.11 & 0.17 \\
           & $R(1,1)^l$ & -3.1 & 8.1 & 0.69 & 0.06 & 3.15 & 0.28 \\

\enddata
\tablecomments{Column 3 ($v_{\rm LSR}$) gives the interstellar gas velocity in the local standard of rest frame.  Column 4 (FWHM) gives the full width at half-maximum of the absorption features.  Columns 5 and 6 show the equivalent width, $W_{\lambda}$, and its $1\sigma$ uncertainty, $\sigma(W_{\lambda})$, respectively.  Columns 7 and 8 give the column density of H$_3^+$ in the state each transition probes, $N(J,K)$, and its uncertainty, $\sigma(N)$, respectively. Values for these parameters in the $\zeta$~Per and X~Per sight lines were previously reported in \citet{indriolo2007}.  The new values for both absorption lines toward $\zeta$~Per and the $R(1,0)$ line toward X~Per are consistent with the previously published results within uncertainties.  However, the new and old results for the $R(1,1)^u$ line toward X~Per are inconsistent.  Upon inspection, we found this to be due to a bad fit to that line during the 2007 analysis.  In all cases, the values published herein should be taken to supersede those from \citet{indriolo2007}.}

\end{deluxetable*}

These observations increase the total number of sight lines with both H$_3^+$ and H$_2$ detections from 2 to 5.  Column densities, $para$-fractions, and excitation temperatures for both species along all 5 sight lines are collected in Table \ref{tblresults}.
%Previously published H$_3^+$ data come from \citet{indriolo2007}, while 
H$_2$ data come from \citet{savage1977} and \citet{rachford2002}.  Uncertainties on all values are $1\sigma$. The excitation temperatures inferred from the $R(1,0)$, $R(1,1)^u$, and $R(1,1)^l$ absorption lines of H$_3^+$ range from $20\leq T$(\hhh)$\leq 46$ K, while those reported for H$_2$ vary from $51\leq T({\rm H}_2)\leq 68$ K.  In 4 sight lines $T_{01}$ is greater than $T({\rm H}_3^+)$ by about 30 K, while for X~Per $T_{01}$ and $T({\rm H}_3^+)$ are consistent within uncertainties.  Still, these observations clearly show that for the same interstellar material along 4 different diffuse molecular cloud sight lines the excitation temperatures derived from H$_3^+$ and H$_2$ do not agree.

\begin{deluxetable*}{lcccccc}
\tablecaption{Target Sight Line Properties \label{tblresults}}
\tablehead{
& & \colhead{$\zeta$ Per\tablenotemark{ab}} & \colhead{X Per\tablenotemark{ac}} & \colhead{HD 154368\tablenotemark{c}} & \colhead{HD 73882\tablenotemark{c}} &\colhead{HD 110432\tablenotemark{cd}}
}
\startdata
& & & H$_3^+$ Results & & & \\
\hline \\
$N(1,1)$  & ($10^{13}$~cm$^{-2}$) & $4.09\pm0.53$  & $3.34\pm0.69$    & $7.43\pm1.24$  & $6.08\pm0.12$  & $3.11\pm0.05$   \\
$N(1,0)$  & ($10^{13}$~cm$^{-2}$) & $2.53\pm0.29$  & $3.29\pm0.45$    & $2.83\pm0.74$  & $2.94\pm0.48$  & $2.11\pm0.17$   \\
$p_3$\tablenotemark{e} &          & $0.62\pm0.04$  & $0.50\pm0.06$    & $0.72\pm0.06$  & $0.67\pm0.04$  & $0.60\pm0.02$   \\
$T({\rm H}_3^+)$  & (K)           & $28\pm4$       & $46^{+21}_{-13}$ & $20\pm4$       & $23\pm3$       & $30\pm2$        \\
\hline \\
& & & H$_2$ Results & & & \\
\hline \\
log$[N(0)]$ & (cm$^{-2}$)         & $20.51\pm0.09$ & $20.76\pm0.03$   & $21.04\pm0.05$ & $20.99\pm0.08$ & $20.40\pm0.03$  \\
log$[N(1)]$ & (cm$^{-2}$)         & $20.18\pm0.09$ & $20.42\pm0.06$   & $20.54\pm0.15$ & $20.50\pm0.07$ & $20.27\pm0.04$  \\
$p_2$\tablenotemark{f} &          & $0.68\pm0.06$  & $0.69\pm0.04$    & $0.76\pm0.07$  & $0.76\pm0.05$  & $0.57\pm0.03$   \\
$T_{01}$  & (K)                   & $58\pm6$       & $57\pm4$         & $51\pm8$       & $51\pm6$       & $68\pm5$        \\
\enddata
\tablecomments{Measured column densities for the lowest lying \ortho\ and \para\ states of H$_2$ and H$_3^+$ are shown for the 5 sight lines with all such data available.  Also shown are the \para-fractions for each species and the rotational temperatures derived from a simple 2-state system analysis.}
\tablenotetext{a}{Updated analysis of H$_3^+$ data originally presented in \citet{indriolo2007}}
\tablenotetext{b}{H$_2$ data from \citet{savage1977}}
\tablenotetext{c}{H$_2$ data from \citet{rachford2002}}
\tablenotetext{d}{May be affected by multiple velocity components \citep{crawford1995}}
\tablenotetext{e}{\phhh\ fraction: $N(1,1)/(N(1,0) + N(1,1))$}
\tablenotetext{f}{\phh\ fraction: $N(1)/(N(0) + N(1))$}
\end{deluxetable*}

\section{\hh\ THERMALIZATION}

Given the large discrepancies between $T_{01}$ inferred from H$_2$ and the excitation temperature of H$_3^+$, it is important to re-examine the assumption that the inferred $T_{01}$ accurately reflects the kinetic temperature of the diffuse molecular clouds.   If this assumption were incorrect, it would be conceivable that H$_3^+$ provides the true (lower) temperature of diffuse molecular clouds.  There are at least 4 reasons this assumption could be invalid: (1) observational errors in the determination of the $J=0$ and $J=1$ column densities of H$_2$; (2) an insufficient frequency of H$^+$ + H$_2$ collisions to achieve steady state; (3) the steady state of this reaction being different from the thermodynamic equilibrium; and (4) errors caused by a varying $J=0:1$ ratio along the line of sight.  In the following subsections, we investigate each of these possibilities in turn.

\subsection{Observational Determination of \hh\ Columns}

The measurement of the column densities of $J=0$ and $J=1$ of H$_2$ is performed by profile fitting to spectra of the Lyman ($A$-$X$) band in the ultraviolet, recorded with Copernicus or FUSE.  The H$_2$ transitions are optically thick, and are completely opaque in the line cores.  Given the difficulties in accurately retrieving column densities from optically thick transitions, one might worry that the inferred $T_{01}$ is contaminated by uncertainties caused by radiative transfer.

According to B.~L.~Rachford (private communication, 2010), the detailed shape of the combined profile of the $J=0$ line and the two $J=1$ lines is quite sensitive to the ratio of the column densities of these two rotational levels, and thus provides a very useful probe of $T_{01}$.  Because multiple vibronic bands of H$_2$, which are known to be relatively free of contamination by stellar lines, are used in the determination of $T_{01}$, it is difficult to envision any systematic errors that could affect the measurements.  The statistical errors in the column density measurements are typically $\sim$0.1 dex, and we can therefore conclude that the ultraviolet measurements provide an accurate and fairly precise measure of the ratio of $N(0)$ to $N(1)$ for H$_2$.

\subsection{Frequency of H$^+$ + \hh\ Reactions}

The $J=0$ and $J=1$ rotational levels of H$_2$ belong to different spin modifications ({\it para} and {\it ortho}, respectively) and are therefore not thermalized by non-reactive collisions or radiative transitions.  Only chemical reactions, in which the protons are exchanged, can affect the nuclear spin modification of H$_2$.\footnote{Strictly speaking, large inhomogeneous magnetic fields, such as found on the surface of paramagnetic catalysts, can also affect the nuclear spin modification of H$_2$, but we assume that such effects are negligible in interstellar conditions.}

Since H$_2$ is formed in a highly exothermic reaction on grain surfaces, its nascent rotational (and spin) distribution is presumed to represent a high temperature \citep{takahashi2001}.  In the high temperature limit, H$_2$ should be formed with an \ortho:\para\ ratio of 3:1.  If an insufficient number of reactive collisions occur between the formation of an H$_2$ molecule and its subsequent destruction (by photodissociation or cosmic-ray ionization), then the average \ortho:\para\ ratio of H$_2$ may lie somewhere between the nascent value (3:1) and the thermalized value (1:2 at 60 K, for example).  This could lead to $T_{01}$ overestimating the true cloud kinetic temperature.

The number of reactive collisions suffered by an average H$_2$ molecule in its lifetime can be expressed as ${\cal N}_{rxn}\equiv\tau_{life}/\tau_{rxn}$, where $\tau_{life}$ is the average lifetime of an H$_2$ molecule and $\tau_{rxn}$ is the average time between reactive collisions.  If ${\cal N}_{rxn}\gg1$, then the \ortho:\para\ ratio of H$_2$ should reflect the steady-state of the reaction in question.

In diffuse molecular clouds, H$_2$ is formed on grains at a rate of $R n_{\rm H} n({\rm H})$, where $R$ is the grain formation rate \citep[typically taken to be about $3\times10^{-17}$~cm$^3$~s$^{-1}$; e.g.][]{spitzer1978,gry2002}, $n_{\rm H} \equiv n({\rm H}) + 2n({\rm H}_2)$ is the total number density of hydrogen nuclei, and $n({\rm H})$ is the number density of atomic hydrogen.  The destruction of H$_2$ is dominated by cosmic-ray ionization and photodissociation (following absorption in the Lyman bands), and has a rate of $(\zeta_2+\Gamma) n({\rm H}_2)$, where $\Gamma$ is the photodissociation rate.  In steady state, these two rates are equal, and we can solve for
$\tau_{life} = (\zeta_2+\Gamma)^{-1} = n({\rm H}_2)/[R n_{\rm H} n({\rm H})]$.  By using the
definition of the local molecular fraction $f^n_{{\rm H}_2} \equiv 2n({\rm H}_2)/n_{\rm H}$ (which we will simply call $f$), we can rewrite this expression
as $\tau_{life} = f/[2R n_{\rm H}(1-f)]$.

Assuming that the reaction of H$^+$ + H$_2$ dominates the interconversion of \ohh\ and \phh (\ortho- and \para-\hh, respectively), we can write $\tau_{rxn} = [k_{ic}n({\rm H}^+)]^{-1}$, where $k_{ic}$ is the rate coefficient for the interconversion reaction.  Substituting into the equation for ${\cal N}_{rxn}$, we find
\begin{displaymath}
{\cal N}_{rxn} = \frac{k_{ic}}{R} \frac{n({\rm H}^+)}{n_{\rm H}} \frac{f}{2(1-f)}.
\end{displaymath}

To estimate the number density of H$^+$, we consider the steady state of its formation and destruction.  H$^+$ is formed by cosmic-ray ionization of H atoms, at a rate of $\zeta_{\rm H}n({\rm H})$, where $\zeta_{\rm H}$ is the cosmic-ray ionization rate of H \citep[$2.3\zeta_{\rm H}\approx1.5\zeta_2$;][]{glassgold1974}.  Given the abundances of various species in diffuse clouds and the rate coefficients for reactions between H$^+$ and such species, chemical models \citep[e.g.][]{woodall2007} predict that H$^+$ is destroyed predominantly by charge transfer to atomic oxygen, with a rate of $k_{ct} n({\rm H}^+) n({\rm O})$.  Equating these rates of formation and destruction and solving for $n({\rm H}^+)$, we find
$n({\rm H}^+)=\zeta_{\rm H}(1-f)/[k_{ct}x({\rm O})]$, where $x({\rm O})\equiv n({\rm O})/n_{\rm H}$.  Finally, substituting this into the expression for ${\cal N}_{rxn}$ gives
\begin{displaymath}
{\cal N}_{rxn} = \frac{k_{ic}}{R} \frac{\zeta_{\rm H}}{k_{ct}} \frac{f}{2n_{\rm H}} \frac{1}{x({\rm O})}.
\end{displaymath}
Adopting values of $k_{ic}=2.2\times10^{-10}$ cm$^3$ s$^{-1}$ \citep{gerlich1990}, $R=3\times10^{-17}$~cm$^3$~s$^{-1}$, $\zeta_2=4\times10^{-16}$~s$^{-1}$ \citep{indriolo2007}, $k_{ct}=7.31\times10^{-10}(T/300)^{0.23}e^{-225.9/T}$~cm$^3$~s$^{-1}$ \citep{woodall2007}, $x(\textrm{O})\approx 3 \times 10^{-4}$ \citep{cartledge2004,jensen2005}, and typical diffuse cloud values of $f = 0.9$\footnote{While the line-of-sight molecular fraction is lower, that quantity integrates over atomic gas not associated with the diffuse molecular cloud. We are therefore using the local molecular fraction typical of diffuse molecular clouds.}  and $n_{\rm H} = 100$ cm$^{-3}$ \citep{snow2006}, we find that at $T\sim70$ K, ${\cal N}_{rxn} \sim 1400$. Thus, the typical H$_2$ molecule will experience over 1000 reactive collisions during its lifetime, and we can safely assume that the initial \ortho:\para\ ratio of H$_2$ is irrelevant; instead, the observed \ortho:\para\ ratio should reflect the steady state of the reactive collisions.

\subsection{Steady State of H$^+$ + \hh\ Reactions}

From a thermodynamic perspective, one would expect that the steady state of the H$^+$ + H$_2$ reaction would represent a thermal distribution of \ohh\ and \phh, if no other processes influence the spin modifications.  This expectation has been confirmed by a phase space theoretical calculation by \citet{gerlich1990}, who found that the \ortho:\para\ ratio could be expressed at low temperatures (30--80 K) by the expression $9.35\exp(-169.4/T)$, quite close to the thermodynamic expectation of $9\exp(-170.4/T)$.  Evidently the nuclear spin selection rules for this chemical reaction, which suppress the \ortho-to-\para\ conversion by a factor of 6, do not significantly impact the final distribution.

\subsection{Line of Sight Integration Effects}

One remaining concern regards the estimation of $T_{01}$ in a diffuse molecular cloud from the column densities of $J=0$ and $J=1$, which are integrated quantities along the line of sight.  If some of the H$_2$ resides in hotter, mostly atomic gas where H$_3^+$ is not abundant, that hot H$_2$ would cause the observed line-of-sight $T_{01}$ to exceed $T_{01}$ in the molecular cloud.  We expect that such contamination would not be a major effect, as H$_2$ is known to self-shield very effectively from the interstellar radiation field; thus, the amount of H$_2$ in primarily atomic (and presumably warmer) gas is likely to be quite small compared to the amount of H$_2$ in the molecular cloud itself.

\begin{figure}
\epsscale{1.0}
\plotone{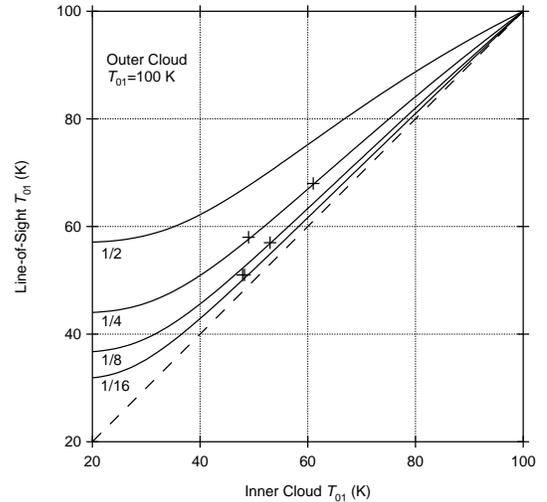}
\caption{The line-of-sight rotational temperature, $T_{01}$, is determined for a cloud containing both a warm and cold component.  The cold, inner component is varied between 10~K and 100~K, while the warm, outer component is set at 100~K.  Different curves show the results for cases where the warm component contains 1/2, 1/4, 1/8, and 1/16 of the total material in the cloud, and are labeled accordingly. Cross hairs mark the inferred inner cloud values of $T_{01}$ given the observed line-of-sight values of $T_{01}$ and estimated fractions of H$_2$ in a 100~K cloud component for the 5 sight lines considered herein.  From left to right the cross hairs mark HD 73882, HD 154368, $\zeta$ Per, X Per, and HD 110432.}
\label{figT01LOS}
\end{figure}

To estimate the magnitude of this effect more quantitatively, we take a simple cloud model with a hotter outer region and cooler inner region.  Assuming that $T_{01}=100$~K in the outer region \citep[based on Copernicus observations of diffuse atomic clouds;][]{jenkins1983}, we varied $T_{01}$ in the inner region between 10~K and 100~K for a set of models where the outer region contained 1/2, 1/4, 1/8, and 1/16 of the material in the cloud.  We then computed the line-of-sight $T_{01}$ that would be derived considering both regions of gas.  The result of this analysis is shown in Figure \ref{figT01LOS}.

Taking the diffuse cloud model of \citet{neufeld2005} with a constant density ($n_{\rm H}=100$~cm$^{-3}$) and standard UV radiation field ($\chi_{\rm UV}=1$) illuminating the cloud from one side, we then determined the amount of H$_2$ expected to be in the outer region for comparison with observed H$_2$ column densities in diffuse clouds.  We define the transition from the outer to inner regions to be at $E(B-V)=0.04$ ($N_{\rm H}\approx2.3\times10^{20}$~cm$^{-2}$), about half of the total color excess (hydrogen column density) found to supply self-shielding effects in H$_2$ \citep{savage1977,gillmon2006}.  Integrating $n({\rm H}_2)$ in the outer region gives $N_{outer}({\rm H}_2)=6\times10^{19}$~cm$^{-2}$.  Because this model effectively only treats one side of a cloud, we compare this value to half of the total H$_2$ column densities reported in Table \ref{tblresults}.  The two extreme cases are HD 110432 and HD 154368, where the outer region accounts for about 1/4 and 1/12 of the total cloud, respectively.  Taking the observed values of $T_{01}$ and using the appropriate curves on Figure 2, we estimate the temperature of the inner cloud region for each of the 5 sight lines considered herein.  The results are marked in Figure 2 as cross hairs, and show that the line-of-sight values of $T_{01}$ overestimate the inferred inner cloud values of $T_{01}$ by only about 5--10 K.  As such, the observed values of $T_{01}$ should be relatively close to the true values of $T_{01}$ in molecular clouds. We therefore assume for the remainder of this paper that the line of sight $T_{01}$ does represent the diffuse molecular cloud kinetic temperature.

\subsection{Summary on H$_2$ Temperature}

From the preceding discussions, we conclude that UV measurements accurately measure the column densities of
$J=0$ and $J=1$ of H$_2$ in diffuse clouds, that the chemical reaction between H$^+$ and H$_2$ occurs $\sim10^3$ times during the life of an average H$_2$, and that the steady state of this chemical reaction leads to an \ortho:\para\
ratio that closely reflects the kinetic temperature of the gas.  Furthermore, we conclude that it is unlikely that the integration along the line of sight introduces significant contamination of the inferred $T_{01}$.  In summary, then, measured values of $T_{01}$ in diffuse molecular clouds should accurately reflect the cloud kinetic temperature.  Consequently, the excitation temperature of H$_3^+$, which is significantly lower than $T_{01}$, must not always reflect the kinetic temperature.

\section{\textit{ORTHO} AND \textit{PARA} H$_3^+$}

Having shown that the temperature discrepancy between $T_{01}$ and $T$(\hhh) in diffuse molecular clouds is real and that $T_{01}$ accurately reflects the cloud kinetic temperature, we now consider the processes related to \hhh\ thermalization in diffuse molecular clouds to examine why $T$(\hhh) might not match the kinetic temperature.

\subsection{Nascent and Thermalized \textit{Para}-H$_3^+$ Fractions} \label{nasctherm}

H$_3^+$ is produced by the reaction
\begin{equation}
\label{formrxn}
\rm{H}_2 + \rm{H}_2^+ \rightarrow \rm{H}_3^+ + H,
\end{equation}
which follows the cosmic-ray ionization of H$_2$ to form H$_2^+$.  The nuclear spin modification of the product H$_3^+$ depends on the nuclear spin modifications of the reactant H$_2$ and H$_2^+$ according to the selection rules given by \cite{oka2004}.  It is most convenient to express the nuclear spin modifications in terms of the \para-fractions, rather than the \ortho:\para\ ratios, so we define
\begin{displaymath}
 p_2 \equiv \frac{n(\textrm{\phh})}{n(\textrm{\phh})+n(\textrm{\ohh})} \label{p2def}
\end{displaymath}
and
\begin{displaymath}
 p_3 \equiv \frac{n(\textrm{\phhh})}{n(\textrm{\phhh})+n(\textrm{\ohhh})} . \label{p3def}
\end{displaymath}

As the cosmic-ray ionization of H$_2$ is not expected to affect the nuclear spin modification, we can further assume that the \para-fraction of H$_2^+$ is also given by $p_2$.  Table \ref{tblh3pform} demonstrates, using these reactant fractions and the nuclear spin branching fractions, that the \para-fraction of newly formed H$_3^+$ is $p_3 = (1/3) + (2/3)p_2$, assuming that the rate for the H$_2^+$ + \hh\ reaction is independent of nuclear spin configuration.

In diffuse molecular clouds, the vast majority of the \hh\ population lies in the lowest \ortho\ and \para\ states, as the temperature of 70 K is well below the energy of the next states (the $J=2$ state lies 510 K above $J=0$, and $J=3$ lies 844 K above $J=1$). Therefore, to derive $p_2$ from astronomical observations we use the formula $p_2 = N(0)/[N(0)+N(1)]$. \hhh\ on the other hand does have energetically accessible \para\ states, as the (2,2) and (2,1) states lie only 151.3 and 249.2 K above the (1,1) ground state. However, the (2,2) and (2,1) states are expected to quickly undergo radiative decay to the (1,1) state at the temperatures and densities of the diffuse molecular clouds \citep{oka2004b}. Furthermore, population in the next \ortho\ state, (3,3) has not been observed in these environments \citep{oka2005}, so the vast majority of \ohhh\ is in the (1,0) state. Consequently, to calculate $p_3$ from the astronomical observations, we use $p_3 = N(1,1)/[N(1,1)+N(1,0)]$.

\begin{deluxetable*}{cccc}
\tablecaption{Nascent Para H$_3^+$ Fraction \label{tblh3pform}}
\tablehead{
\colhead{Reaction} & \colhead{Collision Fraction}  & \colhead{Branching Fraction} & \colhead{\phhh\ Fraction}
}
\startdata
p-H$_2^+$  +  \phh\ & $(p_2)^2$ & 1 & $ p_2^2$ \\
p-H$_2^+$ + \ohh\ & $p_2(1-p_2)$ &  2/3 & $(2/3)(1-p_2)p_2$ \\
o-H$_2^+$ + \phh\ & $(1-p_2)p_2$ &  2/3 & $(2/3)(1-p_2)p_2$ \\
o-H$_2^+$ + \ohh\ & $(1-p_2)^2$ &  1/3 & $(1/3)(1-p_2)^2$ \\
\hline \\
Total & -- & -- & $(1/3) + (2/3)p_2$
\enddata
\tablecomments{This table presents the calculation of the nascent \phhh\ fraction formed in diffuse molecular clouds from the H$_2^+$ + \hh\ reaction, assuming that cosmic ray ionization of \hh\ to form H$_2^+$ does not affect its nuclear spin configuration. The collision fraction represents the fraction of total H$_2^+$ + \hh\ collisions with the specified nuclear spin configurations. The branching fractions are for \phhh\ formation, and are derived from nuclear spin selection rules \citep{quack1977,oka2004}. The final column presents the calculation of the nascent \phhh\ fraction.}
\end{deluxetable*}

\begin{figure}
\epsscale{1.0}
\plotone{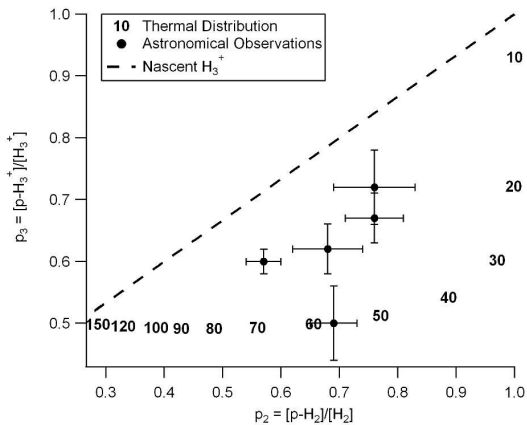}
\caption{The nascent (dashed line) and thermalized (numbers, in K) \phhh\ fraction as a function of the \phh\
fraction.  The circles represent the observations of diffuse molecular clouds summarized in Table \ref{tblresults} with 1$\sigma$ uncertainties.}
\label{p3p2_1}
\end{figure}

Figure \ref{p3p2_1} shows the nascent $p_3$ distribution as a function of $p_2$. This figure also shows the total \para-fraction of a thermalized sample of \hhh\ at various temperatures, calculated using the energy levels $E(J,K)$ from \citet{lindsay2001}.  Also plotted in the figure are the results of the astronomical observations presented in Table \ref{tblresults}. In diffuse molecular clouds, $p_3$ generally appears to lie between the nascent $p_3$ and the thermal $p_3$ values, suggesting an incomplete thermalization of the nuclear spin modifications of H$_3^+$.

As discussed by \citet{oka2004b}, the aforementioned spontaneous emission from the (2,2) and (2,1) states decreases $T$(\hhh) relative to $T_{01}$. They show that for a cloud density of 100 cm$^{-3}$ and $60 \leq T_{01} \leq 120$ K, $T$(\hhh) should fall in the range of 40-50 K, and this accounts for a $\sim40-80$ K difference between $T_{01}$ and $T$(\hhh). However, in the temperature ranges discussed here, the \para-fractions of \hhh\ and \hh\ are nonlinear with respect to these excitation temperatures. In terms of $p_3$, all $T$(\hhh) above about 40 K should have about the same $p_3$, while $p_3$ changes substantially when $T$(\hhh) falls below 40 K, as can be seen in Figure \ref{p3p2_1}. Spontaneous emission will raise the apparent $p_3$ [as derived from the N(1,1):N(1,0) ratio] relative to the thermalized $p_3$ and consequently lower $T$(\hhh) with respect to $T_{01}$. However, Figure \ref{p3p2_1} illustrates that the magnitude of this effect cannot account for the discrepancy observed in the astronomical observations in these environments in terms of the \para-fractions, with the possible exception of X Per.

\subsection{The Reaction of \hhh\ with \hh}

As in the case of H$_2$, the nuclear spin modifications of H$_3^+$ cannot effectively be changed by radiative transitions or by non-reactive collisions; only chemical reactions can do so.  In this case, the reaction \hhh\ + \hh\ $\rightarrow$ (H$_5^+$)* $\rightarrow$ \hh\ + \hhh\ is the most efficient mechanism for interconverting \ohhh\ and \phhh.  When \hhh\ and \hh\ collide, there are three possible reaction outcomes:

\begin{mathletters}
\begin{eqnarray}
{\rm \tilde{H}}_3^+ + {\rm H}_2 \rightarrow & {\rm \tilde{H}}_3^+ + {\rm H}_2 & {\rm (identity),} \label{id} \\
{\rm \tilde{H}}_3^+ + {\rm H}_2 \rightarrow & {\rm H}_2{\rm \tilde{H}}^+ + {\rm \tilde{H}}_2 & {\rm (hop), and} \label{hop} \\
{\rm \tilde{H}}_3^+ + {\rm H}_2 \rightarrow & {\rm H}{\rm \tilde{H}}_2^+ + {\rm H}{\rm \tilde{H}} & {\rm (exchange).} \label{exch}
\end{eqnarray}
\end{mathletters}

In the case of reaction \ref{id}, the nuclear spin configurations of the \hhh\ and \hh\ remain unchanged, while in reactions \ref{hop} and \ref{exch} the configuration may change. However, like reaction \ref{formrxn}, the hop and exchange pathways must obey nuclear spin selection rules \citep{quack1977,oka2004}. For instance, in order for a reaction between \phhh\ and \phh\ to form \ohhh\, the reaction must be an exchange, and \ohh\ must also be formed to conserve the total nuclear spin angular momentum.

A potential energy surface based on high-level \textit{ab initio} calculations is available for the H$_5^+$ system~\citep{xie2005}. Based on the surface stationary points, a barrier of 52.2 cm$^{-1}$ must be overcome for a hop reaction (\ref{hop}) to occur, and a barrier of 1565.9 cm$^{-1}$ for an exchange reaction (\ref{exch}) to occur. The dissociation energy $D_e$ is calculated to be 2903 cm$^{-1}$; therefore (H$_5^+$)$^*$ formed from association of \hhh\ with \hh\ has sufficient energy to overcome these barriers. As such, the complex may undergo many hop and exchange processes over its lifetime, and given sufficient time, the product distribution may approach a statistical limit. As the statistical weights for the hop and exchange reactions are 3 and 6, respectively, the branching ratio $\alpha \equiv k_{\ref{hop}}/k_{\ref{exch}}$ is 0.5 in the statistical limit. Quantum reactive scattering calculations are presently unavailable on this potential energy surface, so experimental data are necessary for determining the value of $\alpha$, as well as its temperature dependence.

The only experimental determination of $\alpha$ for the \hhh\ + \hh\ system was performed by \cite{cordonnier2000}. This study was done by spectroscopically measuring the \ortho-to-\para\ ratio of \hhh\ formed in a discharge of pure \phh\ at 400 K, and under these conditions, the value $\alpha = 2.4$ was derived. No measurements at lower temperatures have been reported for this system, but the isotopically-substituted reaction D$_3^+$ + \hh\ has been studied at a variety of collision energies using an ion trap/guided beam technique~\citep{gerlich1993}. It was observed that $\alpha$ varies substantially with the D$_3^+$-\hh\ collision energy. As this energy decreases, $\alpha$ approaches the statistical value of 0.5, and the value 2.4 is reached at an energy corresponding to the average collision energy at $\sim$440 K, in general agreement with the study by \cite{cordonnier2000}. However, a direct comparison of these results to \hhh\ in the interstellar medium is problematic owing to the endothermicity of the reaction channel and the nonthermal reactant internal state distribution in the experimental measurement.

The final consideration for this reaction is the fraction of reactions that lead to no change in the nuclear spin modification, $S^{id}$. A large value for $S^{id}$ would indicate that nuclear-spin-changing collisions are a small fraction of the total number of \hhh\ + \hh\ collisions, and the thermalization process would be slower than the collision rate. In fact, there is experimental evidence for this, as studies of the \hhh\ + HD $\rightarrow$ H$_2$D$^+$ + \hh\ reaction give a rate coefficient of $3.5 \times 10^{-10}$ cm$^3$ s$^{-1}$ \citep{gerlich2002}, much slower than the Langevin rate coefficient $1.7 \times 10^{-9}$ cm$^3$ s$^{-1}$. These results lead to $S^{id} \sim 0.8$, but it is possible that $S^{id}$ could be different for the purely hydrogenic system, which is thermoneutral rather than exothermic.

\subsection{Steady State \textit{Para}-H$_3^+$ Fraction from H$_3^+$ + H$_2$: ``Bimolecular Reactive Equilibrium''}

After taking into account its chemical physics, does the steady state of the \hhh\ + \hh\ chemical reaction lead to a completely thermalized $p_3$ in the interstellar medium?  To consider this question, we have constructed a simple steady-state model for \ortho\ and \para-\hhh, in terms of nuclear-spin-dependent rate coefficients $k_{xxxx}$ for each potential sub-reaction (e.g., $k_{oppo}$: \ohhh\ + \phh\ $\rightarrow$ \phhh\ + \ohh).  The derivation of this model, which we shall call the ``bimolecular reactive equilibrium'' (BRE) model, is presented in the Appendix, and the resulting expression for $p_3$ is given in the Appendix as equation (\ref{eqp3}).

The rate coefficients themselves were computed using the prescription of \cite{park2007}, which takes into account both the nuclear spin branching fractions as well as energetic considerations at the state-to-state level, using a microcanonical approach.  This work has since been extended by \cite{hugo2009} to deuterated versions of this chemical reaction, and the latter authors report quantitative agreement between the two sets of calculations.  We therefore judge these rate coefficients to be reliable within the context of this theoretical approach.

The Park \& Light code (provided by K. Park, private communication 2009) requires five input parameters: the kinetic temperature, the rotational temperature of \hhh\ and \hh, and the three branching fractions $S^{id}$, $S^{hop}$, and $S^{exch}$. For these calculations, the rotational temperature was held at 10 K in each nuclear spin manifold in order to have the vast majority of \ohhh\ in (1,0) and \phhh\ in (1,1). Therefore, we express the inputs to the model in terms of only three parameters: $T \equiv T_{kin}$, $S^{id}$, and $\alpha$, as $S^{id} + S^{hop} + S^{exch} = 1$ and $\alpha = S^{hop}/S^{exch}$. The code then outputs all of the rate coefficients required in equation (\ref{eqp3}). For a single set of branching fraction values, the rate coefficients were calculated for $10\leq T \leq 160$ K in steps of 10 K, and $p_2$ was set to its thermal value for each calculation.

\begin{figure}
\epsscale{1.0}
\plotone{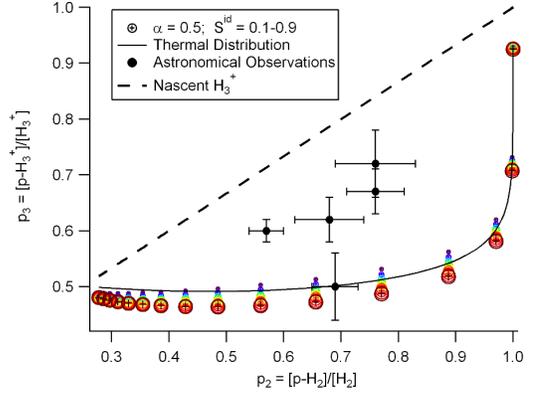}
\caption{BRE calculations of the \para-H$_3^+$ fraction as a function of the \para-H$_2$
fraction, under the influence of the H$_3^+$ + H$_2$ reaction.  The thin solid line represents the thermal limit
(as in Figure \ref{p3p2_1}), and the circled crosses represent the results of our calculations (based on
Park \& Light's model) for $\alpha=0.5$ and various values of $S^{id}$ ranging from 0.1 (small purple) to 0.9 (large brown). Each cluster of crosses represents a calculation at a single temperature, ranging from 10 K (upper right) to 160 K (lower left) in steps of 10 K. Also plotted are the nascent \hhh\ distribution and the astronomical observations.}
\label{p3p2_2}
\end{figure}

\begin{figure}
\epsscale{1.0}
\plotone{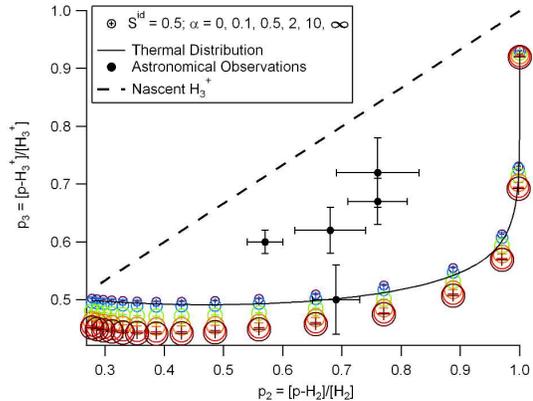}
\caption{Same as Figure \ref{p3p2_2}, except that $S^{id}$ is held at 0.5 and $\alpha$ varies between 0 (small purple) and $\infty$ (large brown).}
\label{p3p2_3}
\end{figure}

Figure \ref{p3p2_2} shows the results of the BRE model for a fixed value of $\alpha=0.5$ for various values
of $S^{id}$ ranging from 0.1-0.9; similarly, Figure \ref{p3p2_3} shows results for fixed $S^{id}=0.5$ and various $\alpha$ ranging from 0-$\infty$.  The results of the calculation are not particularly sensitive to the fraction of collisions that are reactive (as traced by $S^{id}$) or the ratio of the hop to exchange outcomes ($\alpha$). Since in all cases $p_3$ falls near its thermal value, these results suggest that regardless of the values of $\alpha$ or $S^{id}$, the \hhh\ + \hh\ reaction should essentially thermalize the \hhh\ nuclear spin species. This stands in clear contradiction to the reported astronomical observations in diffuse molecular clouds, with the exception of X Per. The discrepancy between $T_{01}$ and $T$(\hhh) cannot be explained by the BRE model, and must then be explained by a lack of equilibration via this chemical reaction.

An interesting aspect of these results is that the steady state $p_3$ at some temperatures is actually {\em below} the value of 0.5 expected based on statistical weights alone (often called the ``high-temperature limit'').  This appears to be a robust result for $\alpha > 0.5$, at least in the range of $S^{id}$ explored here.  This effect may have been observed experimentally in our group's recent measurements of supersonically expanding hydrogen plasmas.  \cite{tom2010} reported $p_3 = 0.491 \pm 0.024$ for a normal hydrogen ($p_2=0.25$) plasma at $\sim80$ K, and \cite{kreckel2010} reported $p_3 = 0.479 \pm 0.02$ in a warmer ($\sim200$ K) normal hydrogen and argon plasma.\footnote{Recent measurements in our laboratory have confirmed, with higher statistical significance, $p_3 < 0.5$ in hollow cathode plasmas containing normal hydrogen.  These results will be reported elsewhere, but it should be noted that $p_3$ in these plasmas may be influenced by three-body collisions due to the higher number densities.} Finally, in work outside our laboratory, \cite{kreckel2007} have reported $p_3 = 0.4$ in a low-temperature ion trap. All of these measurements suggest that it is in fact possible to achieve $p_3 < 0.5$, and lend some evidence to support the calculated results.

\subsection{Steady State \textit{Para}-H$_3^+$ Fraction from H$_3^+$ + H$_2$ and H$_3^+$ + e$^-$}

We now consider whether there are enough reactive collisions within the lifetime of an average H$_3^+$ in
diffuse molecular clouds to bring the spin modifications into BRE.  The destruction of H$_3^+$ in
such clouds is dominated by dissociative recombination (DR) with electrons, and the lifetime is simply $\tau_{life} = (k_{DR} n(e^-))^{-1}$ (the reciprocal of the destruction rate), where $k_{DR}$ is the dissociative
recombination rate coefficient.  The average time between reactive
collisions can be expressed as $\tau_{rxn} = (k_{rc} n(\rm{H}_2))^{-1}$, where $k_{rc}$ is the reactive
collision rate for H$_3^+$ + H$_2$.  The average number of collisions an H$_3^+$ will experience is then ${\cal N}_{rxn} = \tau_{life}/\tau_{rxn} = [k_{rc}/k_{DR}] [n(\textrm{\hh})/n(e^-)]$.

Assuming for the moment that $k_{DR}$ is independent of the nuclear spin modification, we adopt a typical value (for $T\sim70$ K) of $k_{DR}=2 \times 10^{-7}$ cm$^3$ s$^{-1}$ \citep{mccall2004}.  The ratio $n(\textrm{\hh})/n(e^-)$ can be rewritten as $f/2x_e$, where $x_e$ is the electron fraction, typically $1.5 \times 10^{-4}$ assuming charge neutrality and that C$^+$ is the dominant ionic species \citep{cardelli1996,sofia2004}.  If we adopt $f=0.9$, and take $k_{rc}$ to be the full collision rate of H$_3^+$ + H$_2$ \citep[$1.5 \times 10^{-9}$ cm$^3$ s$^{-1}$;][]{adams1987}, we find that ${\cal N}_{rxn} \sim 20$.  However, if we instead adopt the smaller reactive rate coefficient $\sim 3 \times 10^{-10}$ cm$^3$ s$^{-1}$ of \cite{gerlich2002}, we find that ${\cal N}_{rxn} \sim 5$.  With such a small number of collisions in the lifetime of \hhh, $p_3$ may not reach the value predicted by equation \ref{eqp3}.  In the appendix we derive a more complete steady state expression (equation \ref{ssp3}) including the effects of both the H$_3^+$ + H$_2$ reaction and nuclear-spin-dependent DR rates ($k_{e,p}$ and $k_{e,o}$ for \phhh\ and \ohhh).

\begin{figure}
\epsscale{1.0}
\plotone{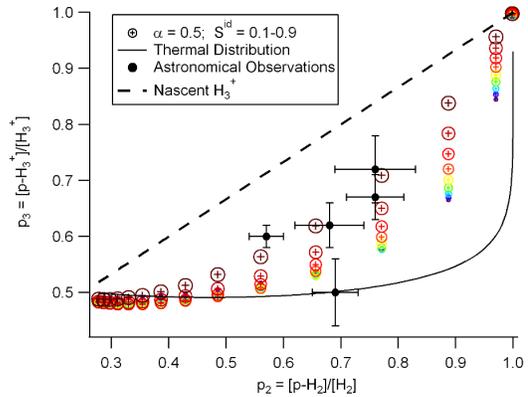}
\caption{Steady state calculations of the \para-H$_3^+$ fraction as a function of the \para-H$_2$
fraction, under the influence of both the H$_3^+$ + H$_2$ reaction and dissociative recombination.
The plotted quantities are analogous to those in Figure \ref{p3p2_2}. In this case, the \ohhh\ and \phhh\ DR rate coefficients $k_{e,o}$ and $k_{e,p}$ are assumed to be equal and taken from \cite{mccall2004}. Each vertical cluster of points represents a calculation at a single temperature, beginning at 160 K in the lower left and decreasing by 10 K each point moving to the right.}
\label{p3p2_4}
\end{figure}

We call this model simply the ``steady state'' model, and we adopt the values $f=0.9$ and $x_e = 1.5 \times 10^{-4}$ as before. Figure \ref{p3p2_4} shows the results of this steady state model if we assume that the DR rate coefficient is the same for both nuclear spin modifications \citep[we have adopted the temperature-dependent value of][]{mccall2004}. In this case, the values of $p_3$ depend quite sensitively on $S^{id}$, as this represents the fraction of \hhh\ + \hh\ collisions that are nonreactive during the relatively short lifetime of \hhh.  Consequently with higher values of $S^{id}$ (larger circles in Figure \ref{p3p2_4}), the \phhh\ fraction in steady state is closer to the nascent fraction.  For $S^{id}=0.9$, which corresponds to a reactive rate coefficient of $k_{rc}=1.9\times10^{-10}$ cm$^{3}$ s$^{-1}$, the calculated $p_3$ are in reasonable agreement with most of the observations. The upper range of the X Per uncertainty is consistent with $S^{id}=0.7$.

\begin{figure}
\epsscale{1.0}
\plotone{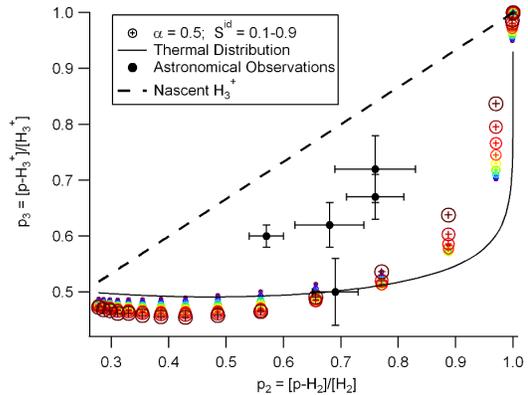}
\caption{Same as Figure \ref{p3p2_4}, except using the spin-dependent dissociative recombination rate coefficients from \cite{santos2007}.}
\label{p3p2_5}
\end{figure}

\begin{figure}
\epsscale{1.0}
\plotone{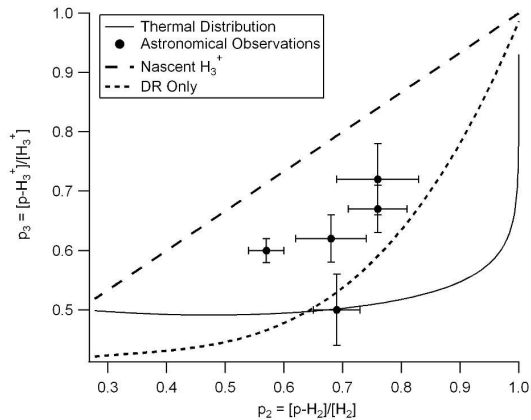}
\caption{Steady state calculations of the \para-H$_3^+$ fraction as a function of the \para-H$_2$
fraction, under the influence of formation and dissociative recombination only.  The solid line shows the thermal limit, and the dotted line represents the results of our calculations, where we have used the spin-dependent dissociative recombination rate coefficients from \cite{santos2007}. Also plotted are the nascent \hhh\ distribution (dashed line) and the astronomical observations.}
\label{p3_noh3h2}
\end{figure}

In Figure \ref{p3p2_5}, we instead consider the calculated DR rate coefficients for \ortho\ and \para-H$_3^+$ presented in \cite{santos2007}.  Their prediction is that \phhh\ is destroyed considerably faster by electrons at low temperatures than \ohhh; consequently, even for large values of $S^{id}$, the steady state $p_3$ does not approach either the nascent or astronomically-observed values. In the absence of the H$_3^+$ + H$_2$ reaction ($S^{id}=1$), $p_3$ would be governed by a steady state determined by the competition between the formation and the spin-dependent DR processes, and this is shown in Figure \ref{p3_noh3h2}.

If the calculated rate coefficients of \cite{santos2007} are correct, it is difficult to explain the observed $p_3$. This is because, with the exception of X Per, the observed $p_3$ are higher than the curve resulting from the steady state of \hhh\ formation and destruction using these DR rate coefficients, and inclusion of the \hhh\ + \hh\ reaction further drives $p_3$ toward the value expected for thermal equilibrium.  Recent storage ring experiments by \cite{tom2009} and \cite{kreckel2005} both saw an increased DR cross-section when \hhh\ is produced from \phh; however, recent imaging results presented in \cite{kreckel2010} suggest that the ions in these experiments have been heated during extraction from the ion sources, and the difference between the \ohhh\ and \phhh\ may therefore have been overestimated.  Further experimental work is clearly needed to pin down the enhancement (if any) in \phhh\ DR, and confirmation of the theoretical predictions would also be quite helpful.

To summarize, according to our models the reaction of \hhh\ with \hh\ is expected to effectively thermalize the nuclear spin configurations of \hhh\ at steady-state, provided that sufficient collisions occur within the lifetime of an \hhh. In diffuse molecular clouds, however, the average number of reactive collisions with \hh\ suffered by an \hhh\ is small, indicating that the formation and destruction rates of the two nuclear spin species should be important. A more complete model which takes these factors into account reaches reasonable agreement with observations in 4 of 5 sight lines provided $S^{id}$ is on the order of 0.9 and \ohhh\ and \phhh\ are destroyed at equal rates owing to DR. Reconciling the observations with the spin-dependent theoretical rates of \cite{santos2007} is difficult, and accurate experimental measurements of the spin-dependent DR rates of \hhh\ at low temperature are needed.

\section{CONCLUSIONS}

While all evidence seems to suggest that $T_{01}$ inferred from ultraviolet spectroscopy of H$_2$ accurately reflects the kinetic temperature of diffuse molecular clouds, the observed excitation temperature of H$_3^+$ is clearly non-thermal in 4 of the 5 measured sight lines.  Based on the microcanonical model of \cite{park2007}, we have constructed a steady state model to predict the \para-H$_3^+$ fraction ($p_3$) if reactive collisions between H$_3^+$ and H$_2$ control the spin modifications of H$_3^+$.  Those results show $p_3$ slightly below the limit expected for full thermalization, and far from the observations.  However, a steady state model that incorporates both the H$_3^+$ + H$_2$ reaction as well as the H$_3^+$ formation (following cosmic-ray ionization) and destruction (by electron recombination) can reproduce the observed $p_3$ if the reactive collision rate is somewhat slow and the dissociative recombination rates for \ortho\ and \para\ H$_3^+$ are comparable.  Our interpretation, given the currently available data, is that H$_3^+$ suffers relatively few spin-changing collisions with H$_2$ in its lifetime, and is thus incompletely equilibrated by this reaction.  The observed \para-H$_3^+$ fraction therefore lies between the nascent fraction and the nearly-thermal fraction that would be reached with sufficient reactive collisions. If our model is correct (and the spin-dependent DR rates of \hhh\ are nearly equal at low temperature), this marks the first determination of the reactive rate coefficient of the \hhh\ + \hh\ reaction, and suggests a value on the order of $10^{-10}$ cm$^{-3}$ s$^{-1}$.

Fully quantum reactive scattering calculations of the H$_3^+$ + H$_2$ reaction would be highly desirable, as they would pin down the state-to-state rate coefficients needed to predict the interstellar \para-H$_3^+$ fraction.  Further experiments and theoretical calculations to elucidate the dependence (if any) of the dissociative recombination on the nuclear spin modification of H$_3^+$ are also urgently needed.  Once the effects of the reactive collisions and dissociative recombination are fully understood, the \para-H$_3^+$ fraction in diffuse molecular clouds can be calculated as a function of the kinetic temperature and the ratio of the molecular fraction to the electron fraction.  This, in turn, suggests that H$_3^+$ may become a useful ``thermometer'' for diffuse molecular clouds with high extinction, where ultraviolet measurements of H$_2$ are not feasible.  However, the calibration of this thermometer will require further experimental and theoretical efforts.

\mbox{}

The authors thank Brian L. Rachford for helpful discussions about the $T_{01}$ determinations, Takeshi Oka for helpful conversations about \ortho\ and \para-H$_3^+$, Kisam Park for providing the code to determine the nuclear-spin-dependent rate coefficients for the \hhh\ + \hh\ reaction, and the anonymous referee for helpful comments.  This work has been supported by NSF grant PHY08-55633.  This work is based in part on observations obtained at the Gemini Observatory, which is operated by the Association of Universities for Research in Astronomy, Inc., under a cooperative agreement with the NSF on behalf of the Gemini partnership: the National Science Foundation (United States), the Science and Technology Facilities Council (United Kingdom), the National Research Council (Canada), CONICYT (Chile), the Australian Research Council (Australia),
Minist\'{e}rio da Ci\^{e}ncia e Tecnologia (Brazil) and Ministerio de Ciencia, Tecnolog\'{i}a e Innovaci\'{o}n Productiva (Argentina).  Gemini/Phoenix spectra were obtained through programs GS-2009B-Q-71 and GS-2010A-Q-60.  This paper is also based in part on observations obtained with the Phoenix infrared spectrograph, developed and operated by the National Optical Astronomy Observatory.

\appendix

\section{REACTIONS AND RATES}

\begin{deluxetable*}{clll}
\tabletypesize{\small}
\tablecaption{Reactions and Rate Equations Used in Models \label{reactions} }
\tablehead{ \colhead{Number} & \colhead{Reaction} & \colhead{Rate} & \colhead{Comments} }
\startdata
1 & \hh\ + CR $\rightarrow$ H$_2^+$ + $e^-$ + CR\' & $\zeta_2$[\hh] & Cosmic ray ionization \\
2 & \hh\ + H$_2^+$ $\rightarrow$ \hhh\ + H & $k_1$[\hh][H$_2^+$] & \hhh\ formation (see Table \ref{formrxn}) \\
3 & $i$-\hhh\ + $j$-\hh\ $\rightarrow$ $m$-\hhh\ + $n$-\hh\ & $k_{ijmn}$[$i$-\hhh][$j$-\hh] & Thermalization reaction for \hhh\ \\
4 & \phhh\ + $e^-$ $\rightarrow$ \hh\ + H (or) 3H & $k_{e,p}$[\phhh][$e^-$] & \phhh\ DR \\
5 & \ohhh\ + $e^-$ $\rightarrow$ \hh\ + H (or) 3H & $k_{e,o}$[\ohhh][$e^-$] & \ohhh\ DR
\enddata
\tablecomments{The branching fractions for \ohhh\ and \phhh\ in reaction 2 are assumed to be given by nuclear spin statistics. In reaction 3, $i$, $j$, $m$, and $n$ represent the nuclear spin configurations of the respective species ($o$ or $p$). Some of these 16 reactions are forbidden by nuclear spin selection rules, and others are not used directly in the derivation because they do not result in a change in the \hhh\ nuclear spin configuration. Square brackets refer to the number density of the species.}
\end{deluxetable*}

In this appendix, we derive the formulas for the bimolecular reactive equilibrium and steady-state \para-H$_3^+$ fractions. We consider 4 processes: cosmic ray ionization of \hh, formation of \hhh, the \hhh\ + \hh\ reaction, and dissociative recombination of \hhh. If all nuclear spin configurations are considered, this gives a total of 28 reactions. The chemical reactions used in the model and their rate expressions are summarized in Table \ref{reactions}. It should be noted that for this section, we employ the chemist's notation of using square brackets to refer to the number density of the respective species.

\section{DERIVATION: BIMOLECULAR REACTIVE EQUILIBRIUM MODEL}

Consider the case that \hhh\ formation and destruction are slow compared with the \hhh\ + \hh\ reaction. We can then ignore formation and destruction processes and  write the rate equation for \phhh\ only in terms of the latter reaction:
\begin{eqnarray}
	\frac{\textrm{d}}{\textrm{d} t}[\textrm{\phhh}] & = & \left\{ \left( k_{oopo} + k_{oopp} \right)[\textrm{\ohh}] + \left( k_{oppo} + k_{oppp} \right) [\textrm{\phh}] \right\}[\textrm{\ohhh}] \nonumber\\
	& & - \left\{ \left( k_{pooo} + k_{poop} \right)[\textrm{\ohh}] + \left( k_{ppoo} + k_{ppop} \right) [\textrm{\phh}] \right\} [\textrm{\phhh}]. \label{dph3dth3h2}
\end{eqnarray}

Assuming steady state, equation \ref{dph3dth3h2} is equal to 0. The right side can then be divided by $[\textrm{\hhh}][\textrm{\hh}]$ in order to express the rate in terms of $p_2$ and $p_3$ as follows:

\begin{eqnarray}
	0 & = & \left\{ \left( k_{oopo} + k_{oopp} \right)(1-p_2) + \left( k_{oppo} + k_{oppp} \right)p_2 \right\}(1-p_3) \nonumber\\
	& & - \left\{ \left( k_{pooo} + k_{poop} \right)(1-p_2) + \left( k_{ppoo} + k_{ppop} \right)p_2 \right\}p_3. \label{h3h2inp3p2ss}
\end{eqnarray}

The resultant equation can be solved for $p_3$:

\begin{equation}
	p_3 = \frac{(k_{oopp}+k_{oopo})(1-p_2) + (k_{oppo}+k_{oppp}) p_2}{(k_{oopp}+k_{oopo}+k_{poop}+k_{pooo})(1-p_2)+(k_{oppo}+k_{oppp}+k_{ppoo}+k_{ppop})p_2 }. \label{p3before}
\end{equation}

Owing to nuclear spin selection rules, the rate coefficients $k_{oppp}$ and $k_{ppop}$ are rigorously 0. Removal of these terms gives the final result:

\begin{equation}
\label{eqp3}
 p_3 = \frac{(k_{oopp}+k_{oopo})(1-p_2) + k_{oppo} p_2}{(k_{oopp}+k_{oopo}+k_{poop}+k_{pooo})(1-p_2)+(k_{oppo}+k_{ppoo})p_2 }. \eqnum{\ref{eqp3}}
\end{equation}

\section{DERIVATION: STEADY STATE MODEL}

Consider now the case in which \hhh\ formation and DR compete effectively with the \hhh\ + \hh\ thermalization reaction. We make the assumption that formation of \phhh\ from H$_2^+$ + \hh\ is governed only by the nuclear spin branching fractions presented in Table \ref{tblh3pform}. Under these conditions, the full rate equation for \phhh is:

\begin{eqnarray}
	\frac{\textrm{d}}{\textrm{d}t}[\textrm{\phhh}] & = & k_1([\textrm{\phh}][\textrm{\textit{p}-H}_2^+] + \frac{2}{3}[\textrm{\phh}][\textrm{\textit{o}-H}_2^+] + \frac{2}{3}[\textrm{\ohh}][\textrm{\textit{p}-H}_2^+] + \frac{1}{3}[\textrm{\ohh}][\textrm{\textit{o}-H}_2^+]) \nonumber \\
	 & & + \left\{ \left( k_{oopo} + k_{oopp} \right)[\textrm{\ohh}] + \left( k_{oppo} + k_{oppp} \right) [\textrm{\phh}] \right\}[\textrm{\ohhh}] \nonumber \\
	& & - \left\{ \left( k_{pooo} + k_{poop} \right)[\textrm{\ohh}] + \left( k_{ppoo} + k_{ppop} \right) [\textrm{\phh}] \right\} [\textrm{\phhh}]  \nonumber \\
	& & - k_{e,p}[e^-][\textrm{\phhh}].
	\label{app2start}
\end{eqnarray}

From Table \ref{tblh3pform}, we can reduce the first line in this equation to $k_1$[\hh][H$_2^+$]$\{(1/3) + (2/3)p_2\}$. We now invoke steady state arguments for all species. For H$_2^+$, $\zeta_2[\textrm{\hh}] = k_1[\textrm{\hh}][\textrm{H}_2^+]$, therefore $k_1[\textrm{H}_2^+]$ can be replaced by $\zeta_2$. Setting the resultant equation equal to zero and dividing by [\hhh][\hh] as before gives:

\begin{eqnarray}
	0 & = & \frac{\zeta_2}{[\textrm{\hhh}]}(\frac{1}{3} + \frac{2}{3}p_2) \nonumber \\
	  &   & + \left\{ \left( k_{oopo} + k_{oopp} \right)(1-p_2) + \left( k_{oppo} + k_{oppp} \right)p_2 \right\}(1-p_3) \nonumber\\
	  &   & - \left\{ \left( k_{pooo} + k_{poop} \right)(1-p_2) + \left( k_{ppoo} + k_{ppop} \right)p_2 \right\}p_3 \nonumber\\
	  &   & - k_{e,p}\frac{[e^-]}{[\textrm{\hh}]}p_3.
	\label{app2_divh3h2}
\end{eqnarray}

This equation can be further simplified by examining the steady state value of [\hhh], which begins with the equation $\zeta_2[\textrm{\hh}] = k_e[\textrm{\hhh}][e^-]$. More specifically, if we include the possibility for different DR rates for \ohhh\ and \phhh, we obtain the equation:

\begin{equation}
	\zeta_2[\textrm{\hh}] = [e^-](k_{e,p}[\textrm{\phhh}] + k_{e,o}[\textrm{\ohhh}]). \label{h3sswithopdr}
\end{equation}

Dividing both sides of equation \ref{h3sswithopdr} by [\hhh][\hh] results in an expression for $\zeta_2$/[\hhh]:

\begin{equation}
	\frac{\zeta_2}{[\textrm{\hhh}]} = \frac{[e^-]}{[\textrm{\hh}]}(k_{e,p}p_3 + k_{e,o}(1-p_3)). \label{zeta2overh3plus}
\end{equation}

Substituting this relation into equation \ref{app2_divh3h2} gives:

\begin{eqnarray}
	0 & = & \frac{[e^-]}{[\textrm{\hh}]}(k_{e,p}p_3 + k_{e,o}(1-p_3))(\frac{1}{3} + \frac{2}{3}p_2) \nonumber \\
	  &   & + \left\{ \left( k_{oopo} + k_{oopp} \right)(1-p_2) + \left( k_{oppo} + k_{oppp} \right)p_2 \right\}(1-p_3) \nonumber\\
	  &   & - \left\{ \left( k_{pooo} + k_{poop} \right)(1-p_2) + \left( k_{ppoo} + k_{ppop} \right)p_2 \right\}p_3 \nonumber\\
	  &   & - k_{e,p}\frac{[e^-]}{[\textrm{\hh}]}p_3.
	\label{app2_sub}
\end{eqnarray}

Solving for $p_3$ and removing the $k_{oppp}$ and $k_{ppop}$ terms yields:

\begin{equation}
 p_3 = \frac{k_{e,o}\frac{[e^-]}{[\textrm{\footnotesize{\hh}}]} \left( \frac{1}{3}+\frac{2}{3}p_2 \right) +(k_{oopp}+k_{oopo})(1-p_2) + k_{oppo} p_2 }{
k_{e,p}\frac{[e^-]}{[\textrm{\footnotesize{\hh}}]} \left( \frac{2}{3}-\frac{2}{3}p_2 \right) +
k_{e,o}\frac{[e^-]}{[\textrm{\footnotesize{\hh}}]} \left( \frac{1}{3}+\frac{2}{3}p_2 \right) +   (k_{oopp}+k_{oopo}+k_{poop}+k_{pooo})(1-p_2)+(k_{oppo}+k_{ppoo})p_2 }. \label{ssp3elech2}
\end{equation}

Finally, the ratio [$e^-$]/[\hh] can be replaced by $2x_e/f$, which results in equation \ref{ssp3}:% from the main text:

\begin{equation}
\label{ssp3}
 p_3 = \frac{k_{e,o}\frac{2x_e}{f} \left( \frac{1}{3}+\frac{2}{3}p_2 \right) +(k_{oopp}+k_{oopo})(1-p_2) + k_{oppo} p_2 }{
k_{e,p}\frac{2x_e}{f} \left( \frac{2}{3}-\frac{2}{3}p_2 \right) +
k_{e,o}\frac{2x_e}{f} \left( \frac{1}{3}+\frac{2}{3}p_2 \right) +   (k_{oopp}+k_{oopo}+k_{poop}+k_{pooo})(1-p_2)+(k_{oppo}+k_{ppoo})p_2 }. \eqnum{\ref{ssp3}}
\end{equation}

% \bibliographystyle{apj}
% \bibliography{h3plus}

\begin{thebibliography}{46}
\expandafter\ifx\csname natexlab\endcsname\relax\def\natexlab#1{#1}\fi

\bibitem[{Adams \& Smith(1987)}]{adams1987}
Adams, N.~G., \& Smith, D. 1987, in IAU Symposium, Vol. 120, Astrochemistry,
  ed. M.~S. Vardya \& S.~P. Tarafdar, 1--18

\bibitem[{Cardelli {et~al.}(1996)Cardelli, Meyer, Jura, \&
  Savage}]{cardelli1996}
Cardelli, J.~A., Meyer, D.~M., Jura, M., \& Savage, B.~D. 1996, \apj, 467, 334

\bibitem[{{Cartledge} {et~al.}(2004){Cartledge}, {Lauroesch}, {Meyer}, \&
  {Sofia}}]{cartledge2004}
{Cartledge}, S.~I.~B., {Lauroesch}, J.~T., {Meyer}, D.~M., \& {Sofia}, U.~J.
  2004, \apj, 613, 1037

\bibitem[{Cordonnier {et~al.}(2000)Cordonnier, Uy, Dickson, Kerr, Zhang, \&
  Oka}]{cordonnier2000}
Cordonnier, M., Uy, D., Dickson, R.~M., Kerr, K.~E., Zhang, Y., \& Oka, T.
  2000, J. Chem. Phys., 113, 3181

\bibitem[{{Crawford}(1995)}]{crawford1995}
{Crawford}, I.~A. 1995, \mnras, 277, 458

\bibitem[{{Dalgarno} {et~al.}(1973){Dalgarno}, {Black}, \&
  {Weisheit}}]{dalgarno1973}
{Dalgarno}, A., {Black}, J.~H., \& {Weisheit}, J.~C. 1973, \aplett, 14, 77

\bibitem[{dos Santos {et~al.}(2007)dos Santos, Kokoouline, \&
  Greene}]{santos2007}
dos Santos, S.~F., Kokoouline, V., \& Greene, C.~H. 2007, J. Chem. Phys., 127,
  124309

\bibitem[{{Gerlich}(1990)}]{gerlich1990}
{Gerlich}, D. 1990, \jcp, 92, 2377

\bibitem[{Gerlich(1993)}]{gerlich1993}
Gerlich, D. 1993, J. Chem. Soc., Faraday Trans., 89, 2199

\bibitem[{{Gerlich} {et~al.}(2002){Gerlich}, {Herbst}, \&
  {Roueff}}]{gerlich2002}
{Gerlich}, D., {Herbst}, E., \& {Roueff}, E. 2002, \planss, 50, 1275

\bibitem[{{Gibb} {et~al.}(2010){Gibb}, {Brittain}, {Rettig}, {Troutman},
  {Simon}, \& {Kulesa}}]{gibb2010}
{Gibb}, E.~L., {Brittain}, S.~D., {Rettig}, T.~W., {Troutman}, M., {Simon}, T.,
  \& {Kulesa}, C. 2010, \apj, 715, 757

\bibitem[{{Gillmon} \& {Shull}(2006)}]{gillmon2006}
{Gillmon}, K., \& {Shull}, J.~M. 2006, \apj, 636, 908

\bibitem[{Glassgold \& Langer(1974)}]{glassgold1974}
Glassgold, A.~E., \& Langer, W.~D. 1974, \apj, 193, 73

\bibitem[{{Goto} {et~al.}(2002){Goto}, {McCall}, {Geballe}, {Usuda},
  {Kobayashi}, {Terada}, \& {Oka}}]{goto2002}
{Goto}, M., {McCall}, B.~J., {Geballe}, T.~R., {Usuda}, T., {Kobayashi}, N.,
  {Terada}, H., \& {Oka}, T. 2002, \pasj, 54, 951

\bibitem[{{Gry} {et~al.}(2002){Gry}, {Boulanger}, {Nehm{\'e}}, {Pineau des
  For{\^e}ts}, {Habart}, \& {Falgarone}}]{gry2002}
{Gry}, C., {Boulanger}, F., {Nehm{\'e}}, C., {Pineau des For{\^e}ts}, G.,
  {Habart}, E., \& {Falgarone}, E. 2002, \aap, 391, 675

\bibitem[{{Hinkle} {et~al.}(2003){Hinkle}, {Blum}, {Joyce}, {Sharp}, {Ridgway},
  {Bouchet}, {van der Bliek}, {Najita}, \& {Winge}}]{hinkle2003}
{Hinkle}, K.~H., {et~al.} 2003, \procspie, 4834, 353

\bibitem[{Hugo {et~al.}(2009)Hugo, Asvany, \& Schlemmer}]{hugo2009}
Hugo, E., Asvany, O., \& Schlemmer, S. 2009, J. Chem. Phys., 130, 164302

\bibitem[{Indriolo {et~al.}(2007)Indriolo, Geballe, Oka, \&
  McCall}]{indriolo2007}
Indriolo, N., Geballe, T.~R., Oka, T., \& McCall, B.~J. 2007, \apj, 671, 1736

\bibitem[{Jenkins {et~al.}(1983)Jenkins, Jura, \& Loewenstein}]{jenkins1983}
Jenkins, E.~B., Jura, M., \& Loewenstein, M. 1983, \apj, 270, 88

\bibitem[{{Jensen} {et~al.}(2005){Jensen}, {Rachford}, \& {Snow}}]{jensen2005}
{Jensen}, A.~G., {Rachford}, B.~L., \& {Snow}, T.~P. 2005, \apj, 619, 891

\bibitem[{{K\"{a}ufl} {et~al.}(2004){K\"{a}ufl}, {Ballester}, {Biereichel},
  {Delabre}, {Donaldson}, {Dorn}, {Fedrigo}, {Finger}, {Fischer}, {Franza},
  {Gojak}, {Huster}, {Jung}, {Lizon}, {Mehrgan}, {Meyer}, {Moorwood}, {Pirard},
  {Paufique}, {Pozna}, {Siebenmorgen}, {Silber}, {Stegmeier}, \&
  {Wegerer}}]{kaufl2004}
{K\"{a}ufl}, H., {et~al.} 2004, \procspie, 5492, 1218

\bibitem[{Kreckel {et~al.}(2005)Kreckel, Motsch, Mikosch, Glosik, Plasil,
  Altevogt, Andrianarijaona, Buhr, Hoffmann, Lammich, Lestinsky, Nevo, Novotny,
  Orlov, Pedersen, Sprenger, Terekhov, Toker, Wester, Gerlich, Schwalm, Wolf,
  \& Zajfman}]{kreckel2005}
Kreckel, H., {et~al.} 2005, Phys. Rev. Lett., 95, 263201

\bibitem[{Kreckel {et~al.}(2007)Kreckel, Petrignani, Berg, Bing, Reinhardt,
  Altevogt, Buhr, Froese, Hoffmann, Jordon-Thaden, Krantz, Lestinsky, Mendes,
  Novotny, Novotny, Pedersen, Orlov, Mikosch, Wester, Plašil, Glosík,
  Schwalm, Zajfman, \& Wolf}]{kreckel2007}
---. 2007, J. Phys.: Conf. Ser., 88, 012064

\bibitem[{Kreckel {et~al.}(2010)Kreckel, Novotn\'y, Crabtree, Buhr, Petrignani,
  Tom, Thomas, Berg, Bing, Grieser, Krantz, Lestinsky, Mendes, Nordhorn,
  Repnow, St\''{u}tzel, Wolf, \& McCall}]{kreckel2010}
---. 2010, \pra, 82, 042715

\bibitem[{Lindsay \& McCall(2001)}]{lindsay2001}
Lindsay, C.~M., \& McCall, B.~J. 2001, J. Mol. Spectrosc., 210, 60

\bibitem[{McCall {et~al.}(1998a)McCall, Geballe, Hinkle, \& Oka}]{mccall1998a}
McCall, B.~J., Geballe, T.~R., Hinkle, K.~H., \& Oka, T. 1998a, Science, 279,
  1910

\bibitem[{{McCall} {et~al.}(1998b){McCall}, {Hinkle}, {Geballe}, \&
  {Oka}}]{mccall1998b}
{McCall}, B.~J., {Hinkle}, K.~H., {Geballe}, T.~R., \& {Oka}, T. 1998b, Faraday
  Discuss., 109, 267

\bibitem[{McCall {et~al.}(2003)McCall, Huneycutt, Saykally, Geballe, Djuric,
  Dunn, Semaniak, Novotny, Al-Khalili, Ehlerding, Hellberg, Kalhori, Neau,
  Thomas, \"Osterdahl, \& Larsson}]{mccall2003}
McCall, B.~J., {et~al.} 2003, Nature, 422, 500

\bibitem[{McCall {et~al.}(2004)McCall, Huneycutt, Saykally, Djuric, Dunn,
  Semaniak, Novotny, Al-Khalili, Ehlerding, Hellberg, Kalhori, Neau, Thomas,
  Paal, Österdahl, \& Larsson}]{mccall2004}
---. 2004, Phys. Rev. A., 70, 052716

\bibitem[{{Neufeld} {et~al.}(2005){Neufeld}, {Wolfire}, \&
  {Schilke}}]{neufeld2005}
{Neufeld}, D.~A., {Wolfire}, M.~G., \& {Schilke}, P. 2005, \apj, 628, 260

\bibitem[{Oka(2004)}]{oka2004}
Oka, T. 2004, J. Mol. Spectrosc., 228, 635

\bibitem[{Oka \& Epp(2004)}]{oka2004b}
Oka, T., \& Epp, E. 2004, \apj, 613, 349

\bibitem[{{Oka} {et~al.}(2005){Oka}, {Geballe}, {Goto}, {Usuda}, \&
  {McCall}}]{oka2005}
{Oka}, T., {Geballe}, T.~R., {Goto}, M., {Usuda}, T., \& {McCall}, B.~J. 2005,
  \apj, 632, 882

\bibitem[{Park \& Light(2007)}]{park2007}
Park, K., \& Light, J.~C. 2007, J. Chem. Phys., 126, 044305

\bibitem[{Quack(1977)}]{quack1977}
Quack, M. 1977, Mol. Phys., 34, 477

\bibitem[{{Rachford} {et~al.}(2002){Rachford}, {Snow}, {Tumlinson}, {Shull},
  {Blair}, {Ferlet}, {Friedman}, {Gry}, {Jenkins}, {Morton}, {Savage},
  {Sonnentrucker}, {Vidal-Madjar}, {Welty}, \& {York}}]{rachford2002}
{Rachford}, B.~L., {et~al.} 2002, \apj, 577, 221

\bibitem[{{Rachford} {et~al.}(2009){Rachford}, {Snow}, {Destree}, {Ross},
  {Ferlet}, {Friedman}, {Gry}, {Jenkins}, {Morton}, {Savage}, {Shull},
  {Sonnentrucker}, {Tumlinson}, {Vidal-Madjar}, {Welty}, \&
  {York}}]{rachford2009}
---. 2009, \apjs, 180, 125

\bibitem[{Savage {et~al.}(1977)Savage, Bohlin, Drake, \& Budich}]{savage1977}
Savage, B.~D., Bohlin, R.~C., Drake, J.~F., \& Budich, W. 1977, \apj, 216, 291

\bibitem[{Snow \& McCall(2006)}]{snow2006}
Snow, T.~P., \& McCall, B.~J. 2006, \araa, 44, 367

\bibitem[{{Sofia} {et~al.}(2004){Sofia}, {Lauroesch}, {Meyer}, \&
  {Cartledge}}]{sofia2004}
{Sofia}, U.~J., {Lauroesch}, J.~T., {Meyer}, D.~M., \& {Cartledge}, S.~I.~B.
  2004, \apj, 605, 272

\bibitem[{{Spitzer}(1978)}]{spitzer1978}
{Spitzer}, Jr., L. 1978, Physical Processes in the Interstellar Medium (New
  York Wiley-Interscience)

\bibitem[{{Takahashi}(2001)}]{takahashi2001}
{Takahashi}, J. 2001, \apj, 561, 254

\bibitem[{Tom {et~al.}(2010)Tom, Mills, Wiczer, Crabtree, \& McCall}]{tom2010}
Tom, B.~A., Mills, A.~A., Wiczer, M.~B., Crabtree, K.~N., \& McCall, B.~J.
  2010, \jcp, 132, 081103

\bibitem[{Tom {et~al.}(2009)Tom, Zhaunerchyk, Wiczer, Mills, Crabtree,
  Kaminska, Geppert, Hamberg, af~Ugglas, Vigren, van~der Zande, Larsson,
  Thomas, \& McCall}]{tom2009}
Tom, B.~A., {et~al.} 2009, \jcp, 130, 031101

\bibitem[{{Woodall} {et~al.}(2007){Woodall}, {Ag{\'u}ndez}, {Markwick-Kemper},
  \& {Millar}}]{woodall2007}
{Woodall}, J., {Ag{\'u}ndez}, M., {Markwick-Kemper}, A.~J., \& {Millar}, T.~J.
  2007, \aap, 466, 1197

\bibitem[{{Xie} {et~al.}(2005){Xie}, {Braams}, \& {Bowman}}]{xie2005}
{Xie}, Z., {Braams}, B.~J., \& {Bowman}, J.~M. 2005, \jcp, 122, 224307

\end{thebibliography}

%%%%%%%%%%%%%%%%%%%%%%%%%%%%%%%%%%%%%%%%%figures%%%%%%%%%%%%%%%%%%%%%%%%%%%%%%%%%%%%%%%%%%

%%%%%%%%%%%%%%%%%%%%%%%%%%%%%%%%%%%%%%%%%tables%%%%%%%%%%%%%%%%%%%%%%%%%%%%%%%%%%%%%%%%%%%

\end{document}